\documentclass[aps,prl,twocolumn,groupedaddress,showpacs,floats]{revtex4}
\usepackage{graphics}
\usepackage{amssymb}
\begin{document}

\title{Weak-Field Thermal Hall Conductivity in the Mixed State of
$d$-Wave Superconductors}
\author{Adam C. Durst, Ashvin Vishwanath, and Patrick A. Lee}
\affiliation{Department of Physics, Massachusetts Institute of Technology,
Cambridge, Massachusetts 02139}
\date{June 6, 2002}

\begin{abstract}
Thermal transport in the mixed state of a $d$-wave superconductor
is considered within the weak-field regime.  We express the
thermal conductivity, $\kappa_{xx}$, and the thermal Hall conductivity,
$\kappa_{xy}$, in terms of the cross section for quasiparticle scattering
from a single vortex.  Solving for the cross section
(neglecting the Berry phase contribution and the anisotropy of the
gap nodes), we obtain $\kappa_{xx}(H,T)$ and $\kappa_{xy}(H,T)$
in surprisingly good agreement with the qualitative features of the
experimental results for YBa$_{2}$Cu$_{3}$O$_{6.99}$.
In particular, we show that the simple, yet previously unexpected,
weak-field behavior, $\kappa_{xy}(H,T) \sim T\sqrt{H}$,
is that of thermally-excited nodal quasiparticles, scattering primarily
from impurities, with a small skew component provided by vortex scattering.
\end{abstract}


\maketitle

Thermal Hall conductivity provides the most direct measure of low temperature
quasiparticle transport in a $d$-wave superconductor.  Since quasiparticles
are part electron and part hole, their energy is well defined but their
charge is not.  Thus, it is thermal current that follows quasiparticle
current.  The longitudinal
thermal conductivity, $\kappa_{xx}$, has both an electronic and a
phononic contribution.  However, the thermal Hall conductivity, $\kappa_{xy}$,
induced by a perpendicular magnetic field (the Righi-Leduc effect),
is purely electronic in origin and the direct consequence of a transverse
quasiparticle current.

Over the past few years, much progress has been made in measuring
the thermal Hall conductivity of the cuprate superconductors in
the mixed (vortex) state, $H_{c1} < H < H_{c2}$
\cite{kri99,ong99,oca01,zha01}.  Most recently, Ong and co-workers
\cite{zha01} measured $\kappa_{xy}$ in high-purity single crystals
of slightly overdoped ($T_{c}=89~K$) YBa$_{2}$Cu$_{3}$O$_{6.99}$
(YBCO). Their data indicates that, for magnetic fields up to
14~Tesla and temperatures between 15~K and 28~K,
$\kappa_{xy}/T^{2}$ is only a function of the ratio $\sqrt{H}/T$.
This is in agreement with the scaling theory proposed by Simon and
Lee \cite{sim97} which predicts that, for nodal quasiparticles
with a Dirac-like dispersion, $\kappa_{xy}(H,T) \sim T^{2}
F_{xy}(\sqrt{H}/\gamma T)$ where $\gamma=(k_{B}/v_{f})\sqrt{c /
\hbar e}$ and $F_{xy}(x)$ is a general scaling function.
Furthermore, the experiments show that for $\sqrt{H} \ll \gamma T$
the measured scaling function has
the surprisingly simple form, $F_{xy}(x) \sim x$, such that
$\kappa_{xy}(H,T) = C_{0} T \sqrt{H}$ where $C_{0}$ is a constant.
For larger magnetic fields, the measured curves peak and then
decrease. It is unusual to see $\sqrt{H}$ (rather than $H$) in a
Hall response and this interesting result was theoretically
unexpected.  Yet such a simple functional form must have a simple
explanation.  In this Letter, we seek to provide it.

The key to this problem is that we are dealing
with the {\it high} temperature regime of low $T$
quasiparticle transport.  By low $T$ quasiparticle transport,
we mean that quasiparticles are excited only in the vicinity of gap nodes
and inelastic scattering can be neglected.  (Quasiparticle dispersion is
therefore given by the anisotropic Dirac spectrum,
$E=(v_{f}^{2}k_{1}^{2}+v_{2}^{2}k_{2}^{2})^{1/2}$,
where $v_{f}$ is the Fermi velocity, $v_{2}$ is the slope of the gap, and
$k_{1}$ and $k_{2}$ are defined locally about each node.  With our choice
of axes, gap nodes are located at $\pm p_{F}\hat{\bf x}$
and $\pm p_{F}\hat{\bf y}$ in momentum space.)
However, the experiments involve temperatures large compared to the
impurity scattering rate and the vortex scattering rate.
As a result, the quasiparticles
responsible for transport are thermally generated rather than
impurity-induced \cite{lee93,gra96,dur00} or
magnetic field-induced \cite{vol93,kop96}.
(Note that thermal transport in the opposite, low $T$, regime has been
discussed frequently in the recent literature
\cite{fra99,vek99,ye01,vis01,vaf01a}.)
This high $T$ regime is relatively simple.
To understand the thermal conductivity, we need only understand how
the thermally excited quasiparticles scatter from impurities and
magnetic vortices.

In the presence of a magnetic field ($H_{c1} < H \ll H_{c2}$),
vortices penetrate the sample (a 2D CuO$_{2}$ layer).
They are distributed randomly, pinned to local defects.  The cuprates
are extreme type II superconductors in which the coherence length, $\xi$,
is much smaller than the penetration depth, $\lambda$.  As a result,
while the vortex cores may be well separated, the magnetic field
profiles overlap significantly such that there is little variation
in the magnetic field across the sample.  We therefore adopt
the extreme type II limit of $\xi \rightarrow 0$ and
$\lambda \rightarrow \infty$ and take the magnetic field to
be constant, ${\bf H} = H \hat{\bf z}$.  In this limit, there are
only two remaining length scales.  The first, $1/k$, is set by
the temperature such that $k \equiv E / \hbar v_{f} = k_{B} T / \hbar v_{f}$.
The second length scale,
$R$, is half of the average distance between vortices.  With one flux
quantum per vortex, $H \pi R^{2} = \Phi_{0} = hc/2e$, so we define
$R \equiv \sqrt{\hbar c / e H}$.  In terms of $R$, we can define the
(2D) density of vortices to be $n_{v} = H/\Phi_{0} = 1/\pi R^{2}$.
The ratio of the two length scales yields
$kR = \gamma T / \sqrt{H}$,
which is the inverse of the argument of the scaling functions.

We consider nodal quasiparticles carrying a heat current in
response to a thermal gradient in the $x$-direction.
Defining a mean free path, $\ell$,
we can express $\kappa_{xx}$ in terms of the electronic specific
heat, $C_{v}$, via $\kappa_{xx} = v \ell C_{v} / 2$
where $v$ is the average quasiparticle velocity \cite{kappafootnote}.
Specific heat in the mixed state has been calculated by
Kopnin and Volovik \cite{kop96}.
For magnetic fields small compared to the temperature, they find
$C_{v} \sim T^{2} [1 + \mathcal{O}((\sqrt{H}/\gamma T)^{2})]$.
The second term, a measure of the magnetic field contribution to
quasiparticle generation, can be neglected in the regime of interest.
Doing so and defining a thermal Hall angle,
$\tan\theta_{H} \equiv \kappa_{xy}/\kappa_{xx}$,
yields a simple form for the thermal conductivity:
$\kappa_{xx} / T = \alpha_{0} k\ell$ and
$\kappa_{xy} / T = \alpha_{0} k\ell \tan\theta_{H}$,
where $\alpha_{0} \approx 1.72 k_{B}^{2} v / \hbar v_{2}$ times the
stacking density of CuO$_{2}$ planes.

The mean free path has contributions from both
impurity scattering and vortex scattering.  For small impurity
densities and dilute vortices, we expect these
to be relatively independent.  Thus, via Matthiessen's rule, we write
$1/\ell = 1/\ell_{0} + 1/\ell_{v}$ where $\ell_{0}$ and $\ell_{v}$
are the contributions from impurities and vortices respectively.
Since $1/\ell_{v}$ vanishes for $H=0$, $\ell_{0}$ can be found
empirically from $\kappa_{e}(T)$, the electronic part of $\kappa_{xx}(H=0)$.
We define $A \equiv k\ell_{0} = \kappa_{e}(T) / \alpha_{0} T$.
As argued by Simon and Lee \cite{sim97}, we expect scaling even in
the presence of disorder as long as the impurity scattering does not
yield an additional length scale (as for $\delta$-correlated disorder).
If so, then $A$ must be $T$-independent.
(Note that it is a different constant from that obtained in
the universal limit \cite{dur00}.)
Zero-field measurements of $\kappa_{xx}$ in YBCO \cite{zhang1,ong99}
show that this is realized experimentally for $T < 30~\mbox{K}$,
which is precisely the temperature range
over which the $\kappa_{xy}$ data obeys scaling.

Now consider the vortex contribution.  Superflow circulates
about the vortices, falling off like $1/r$
near the center of each vortex and cancelling with superflow of
opposite direction (circulating about neighboring vortices) in
the regions between vortices.  We therefore
attribute to each vortex the circulating superflow in the area
that surrounds it.  Since the vortex scattering should depend
primarily on the average vortex density and be relatively
insensitive to higher order details of the vortex configuration,
we can approximate the exact superflow distribution
by cutting off the superflow about each vortex at a distance $R$
from its center.  In this manner, we define an effective single
vortex for which we have approximately accounted for the influence
of neighboring vortices.  If we assume that scattering events
from such vortices are otherwise uncorrelated, then we can
express $\ell_{v}$ in terms of the single vortex transport cross section,
$\sigma_{\parallel}$, and the density of vortices, $n_{v}$.
In this approximation, the transport scattering rate is
$1/\tau_{v} = n_{v} v \sigma_{\parallel}$ and, since $\ell_{v} = v \tau_{v}$,
we find
$\ell_{v} = 1 / n_{v} \sigma_{\parallel} = \pi R^{2} / \sigma_{\parallel}$.
Since vortices are endowed with a circulation,
the vortex scattering cross section can have
a small skew component, $\sigma_{\perp}$.
Thus, when a quasiparticle encounters a vortex, it has a
slightly greater probability of scattering to one side than
the other and thereby contributes to $J_{y}$.  This process
repeats with each successive vortex until the quasiparticle has
travelled a distance equal to its mean free path.  Therefore,
given a heat current, $J_{x}$, we can express the transverse
heat current as $J_{y} = J_{x} n_{v} \sigma_{\perp} \ell$.  The thermal
Hall angle is then given by
$\tan\theta_{H} = \kappa_{xy} / \kappa_{xx}
= - J_{y} / J_{x} = - \sigma_{\perp} \ell / \pi R^{2}$
where we used $\kappa_{xy}=-\kappa_{yx}$.

Combining our results to this point, we find that
$\kappa_{xx}$ and $\kappa_{xy}$ can be expressed in terms of the
single vortex scattering cross section via
\begin{equation}
\frac{\kappa_{xx}}{\alpha_{0} T} \equiv F_{xx}(x)
= \frac{1}{\frac{1}{A}
+ \frac{1}{\pi} x^{2} f_{\parallel}\left(\frac{1}{x}\right)}
\label{eq:KxxfromCSx}
\end{equation}
\begin{equation}
\frac{\kappa_{xy}}{\frac{\alpha_{0} k_{B}}{2E_{F}} T^{2}}
\equiv F_{xy}(x)
= \frac{\frac{1}{\pi} x^{2} f_{\perp}\left(\frac{1}{x}\right)}
{\left(\frac{1}{A}
+ \frac{1}{\pi} x^{2} f_{\parallel}\left(\frac{1}{x}\right)\right)^{2}}
\label{eq:KxyfromCSx}
\end{equation}
where $x \equiv 1/kR = \sqrt{H}/\gamma T$, $E_{F} \equiv v_{f}p_{F}/2$,
and we have defined dimensionless functions of $kR=1/x$ such that
$k\sigma_{\parallel} \equiv f_{\parallel}(kR)$ and
$k\sigma_{\perp} \equiv -(k/p_{F})f_{\perp}(kR)$.
The extra factor of $k/p_{F}$ in the definition of $f_{\perp}$ reflects
the fact that $\kappa_{xy}$ is small by a factor
of $k_{B}T/E_{F}$ \cite{sim97}.

Next, we calculate $f_{\parallel}(kR)$ and
$f_{\perp}(kR)$ by considering the quantum mechanical scattering
of a quasiparticle from a single vortex.
Our calculation, the details of which will be presented elsewhere
\cite{dur02}, is similar in spirit to that conducted by Cleary
\cite{cle68} for the case of an $s$-wave superconductor.
We consider the Bogoliubov-de Gennes (BdG) equation
for a $d$-wave superconductor in the presence of a single vortex.
We apply a singular gauge transformation
that simplifies the BdG Hamiltonian at the cost of imposing antiperiodic
boundary conditions which
require that our wave function change sign with each trip around the vortex.
We further simplify by shifting the origin of momentum space to the location
of one of the gap nodes and neglecting scattering from one node to another.
(This is physically reasonable since the superflow from which quasiparticles
scatter is smooth on the scale of $1/p_{F}$.)
The resulting problem is one of an
(anisotropic) Dirac fermion scattering from an effective non-central
potential (due to the superflow) in the presence of antiperiodic
boundary conditions and small, yet important, curvature terms in the
Hamiltonian.

Quasiparticles interact with vortices via the superflow as well as
the Berry phase factor of (-1) acquired upon circling a vortex.
This phase is encoded in the antiperiodic boundary conditions imposed on
quasiparticles in our chosen gauge.  We make the following
approximations.  First, we neglect the Berry phase effect and instead
adopt periodic boundary conditions for the quasiparticle, which is
equivalent to considering the case of an $hc/e$ (double) vortex.
To justify this, we note that the Berry phase only affects quasiparticle
trajectories that lie within the thermal deBroglie wavelength of the vortex
core, such that two paths that pass on either side of the core can
interfere.  Thus, any additional contribution to the cross section can at best
equal the deBroglie wavelength.  Since this wavelength is much smaller than
the spacing between vortices (for $\sqrt{H} \ll \gamma T$), the Berry phase
contribution may change the size of the cross section, but should not change
its dependence on magnetic field.
Second, we assume an isotropic Dirac dispersion in the linearized Hamiltonian,
which allows us to more easily work with angular momentum eigenstates
and simplifies the calculation.  This is clearly an approximation for the
cuprates where $v_{f}$ exceeds $v_{2}$ by a factor of 10 to 20
\cite{chi00}.  Third, we approximate the effect of neighboring
vortices by cutting off the superflow distribution about our
single vortex at a distance $R$ from its center.  By construction,
the flux through this circle is exactly one ($hc/2e$) flux quantum.
The resulting superfluid momentum (superflow) is
${\bf P}_{s}({\bf r}) = (\hbar/2)(1/r - r/R^{2})\theta(R-r)\hat{\bf \phi}$.
The BdG Hamiltonian (for quasiparticles about the node at $+p_{F}\hat{\bf x}$)
is then given by the sum of a linearized (Dirac) part,
$H_{D} = v_{f} [ \tau_{3} p_{x} + \tau_{1} p_{y} + P_{sx} ]$,
and a quadratic (curvature) part,
$H_{C} = (v_{f}/2p_{F}) [ \tau_{3} (p^{2} + P_{s}^{2})
+ 2 {\bf P}_{s} \cdot {\bf p} + \tau_{1} (2 p_{x} p_{y}) ]$,
which is small sufficiently far from the vortex center ($r>1/p_{F}$).
Since we cutoff our model at the scale of the vortex core
($\xi \sim 10/p_{F}$), curvature terms can be considered perturbatively.
Finally, we select a reasonable core size
and model the vortex core as a region
with vanishing superflow; which is the best we can do in the absence
of further experimental input.  We now have a well-defined scattering
problem, which is first solved considering the linearized Hamiltonian,
and then perturbed to first order in the curvature terms.  (Note, if
curvature terms are neglected completely, there is no skew
scattering \cite{sim97,ye01,vis01}.)  Due to these approximations,
we cannot expect our results to be quantitative.  Yet,
the qualitative agreement with experiment
is surprisingly good, which indicates that we have
retained the essential physics.

Note that $H_{D}$ includes an effective
non-central potential, $V=-v_{f}P_{s}(r)\sin\phi$.  This
mixes angular momentum eigenstates and requires
that we solve for all of them simultaneously.  We do so
numerically, including the contributions of up to 46 angular momenta.
Given the eigenstates both inside ($r<R$) and outside ($r>R$) the vortex,
we apply boundary conditions at the origin, match solutions at the
vortex edge ($r=R$), and construct a wave function composed of an incident
plane wave and an outgoing radial wave.  The angular prefactor of the
radial piece yields the differential cross section.  Summing over
contributions from each of the four nodes and
integrating over scattering angles, we obtain the total cross section,
$\sigma = \int d\varphi\, (d\sigma/d\varphi)$,
the transport cross section,
$\sigma_{\parallel} = \int d\varphi\, (1-\cos\varphi)\, (d\sigma/d\varphi)$,
and the skew cross section,
$\sigma_{\perp} = \int d\varphi\, \sin\varphi\, (d\sigma/d\varphi)$.

\begin{figure}
\centerline{\resizebox{3.0in}{!}{\includegraphics{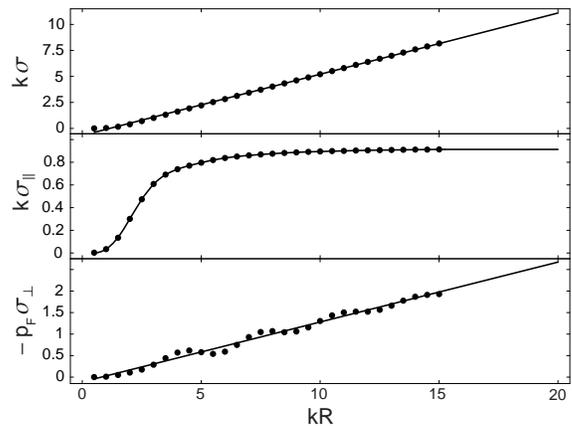}}}
\caption{Calculated total, transport, and skew cross sections as a
function of $kR$ from 0.5 to 15.
For $\sigma$ and $\sigma_{\perp}$, solid lines are fits to
straight lines with small negative intercepts.  For $\sigma_{\parallel}$,
the solid line is an interpolation of the numerical data in which
the constant large-$kR$ behavior is extrapolated to larger $kR$.
The oscillating component of the $\sigma_{\perp}$ data is believed to
be a numerical artifact.}
\label{fig:CSdata}
\end{figure}

Results for a range of intervortex distances, $kR=\gamma T / \sqrt{H}$,
from 0.5 to 15, are plotted in Fig.~\ref{fig:CSdata}.  Note that
while $\sigma$ and $\sigma_{\parallel}$ are plotted in units of
$1/k$, $\sigma_{\perp}$ is plotted in units of $-1/p_{F}$.  This
reflects the fact that the skew cross section, induced by the
curvature terms of the Hamiltonian, is small by a factor of $k/p_{F}$.
The minus sign indicates that the quasiparticles get deflected to the
right, just as an electron would in response to the Lorentz force.
For large $kR$, $k\sigma$ and $-p_{F}\sigma_{\perp}$ increase linearly
with $kR$ while $k\sigma_{\parallel}$ saturates to a constant value.

The form of $\sigma$ and $\sigma_{\parallel}$ can be understood
in terms of a simple Born-limit calculation \cite{dur02}.  The 2D Fourier
transform of $1/r$ yields (approx) $1/q$ and squares to yield
$d\sigma/d\varphi \sim k/q^{2}$ where
$q^{2}=|{\bf k}-{\bf k}^{\prime}|^{2}=2k^{2}(1-\cos\varphi)$.
Since $1/(1-\cos\varphi)$ diverges for small angles, angular integration
is dominated by the small angle cutoff at $q \approx 1/R$
and yields $\sigma \sim R$.  By constrast, since the extra factor of
$(1-\cos\varphi)$ in the definition of $\sigma_{\parallel}$ precisely
cancels this divergence, we obtain $\sigma_{\parallel} \sim 1/k$.
A nonzero $\sigma_{\perp}$, however, can only be obtained by
going beyond the Born limit.

\begin{figure}
\centerline{\resizebox{3.375in}{!}{\includegraphics{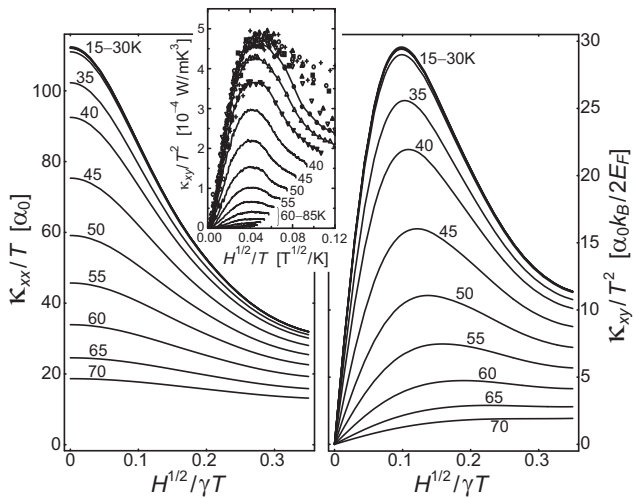}}}
\caption{Calculated longitudinal thermal conductivity (left) and thermal
Hall conductivity (right) plotted in scaling form (in units defined in
the text).  The inset is the measured thermal Hall conductivity
(in SI units) obtained by Ong and co-workers for
YBa$_{2}$Cu$_{3}$O$_{6.99}$ and reproduced here from
Ref.~\onlinecite{zha01}.}
\label{fig:devfromscaling}
\end{figure}

We can now use the fits to these numerical results as the input to
Eqs.~(\ref{eq:KxxfromCSx}) and (\ref{eq:KxyfromCSx}).
The other input, $A \equiv \kappa_{e}(T)/\alpha_{0} T$, is obtained
empirically from the measured zero-field thermal conductivity in
YBCO (extrapolated for the lowest $T$) \cite{zhang1}.
This quantity is a $T$-independent constant for $T < 30$~K but
decreases for larger $T$ where inelastic scattering becomes significant.
Our calculated thermal conductivities are plotted (in scaling form)
in Fig.~\ref{fig:devfromscaling}.  In each plot
we show 15 curves for $T$ ranging from 15~K to 70~K.
The low $T$ curves (for which $A\approx\mbox{const}$) satisfy
scaling and therefore lie nearly on top of each other.
At higher $T$, the curves deviate from scaling,
presumably due to the onset of inelastic scattering.  Both the functional
form of the scaling curves and the manner in which scaling is violated
agree qualitatively with the mixed state thermal conductivity data
measured in YBCO by Ong and co-workers \cite{zha01}.

The form of our scaling curves can be understood as follows.
In the small $H$ regime, the mean free path
is dominated by impurity scattering ($\ell_{0} \ll \ell_{v}$).
Since $x=1/kR=\sqrt{H}/\gamma T$ is small, the
cross sections take on their simple large-$kR$ form,
$f_{\parallel}(1/x) \approx c_{\parallel}$ and
$f_{\perp}(1/x) \approx c_{\perp}/x$, where $c_{\parallel}$ and
$c_{\perp}$ are constants. Therefore, the scaling functions are
$F_{xx}(x) = A$ and $F_{xy}(x) = c_{\perp} A^{2} x / \pi$
and we find that the weak-field thermal Hall conductivity is
\begin{equation}
\kappa_{xy}(H,T) = C_{0} T \sqrt{H}
\label{eq:Kxysimple}
\end{equation}
where $C_{0} \equiv \alpha_{0} k_{B} c_{\perp} A^{2} / 2\pi \gamma E_{F}$.
(For reasonable parameter values, this yields
$C_{0} = 7.2 \times 10^{-3} {\rm W}/{\rm mK}^{2}\sqrt{\rm Tesla}$
which is only a factor of 2 smaller than the experimental value
reported in Ref.~\onlinecite{zha01}.)
As $H$ grows, vortex scattering increases,
increasing the skew scattering while reducing $\ell$
(and therefore $\kappa_{xx}$).  Near the point where
$\ell_{0} \approx \ell_{v}$, the competition between the decreasing
mean free path and the increasing skew scattering results in a peak
for the $\kappa_{xy}$ scaling curve.  If the impurity scattering is
sufficiently weak,
this peak can occur for small enough $H$ that the
cross sections still take their simple large-$kR$ forms.
Thus, for sufficiently clean samples, at temperatures where scaling
is satisfied, it follows from
Eq.~(\ref{eq:KxyfromCSx}) that
the peak height and location should be
proportional to $A^{3/2}$ and $A^{-1/2}$ respectively.
Recall, as well, that $C_{0} \sim A^{2}$.
These are predictions that can be checked experimentally
by comparing results for samples of differing purity.
Once vortex scattering dominates the mean free path
($\ell_{0} \gg \ell_{v}$), $\kappa_{xy}$ decreases with increasing $H$.
As we push this model toward the strong-field (large $x$)
regime, the transport cross section decreases from its large-$kR$ value,
increasing $\ell$ and causing a leveling out and eventual
upturn in the scaling curves for both $\kappa_{xx}$ and $\kappa_{xy}$.
However, in this strong-field regime ($\sqrt{H} \gtrsim \gamma T$),
there is a magnetic
field contribution to the quasiparticle density of states and so our
picture of thermally-excited quasiparticles scattering from dilute
vortices ceases to be valid.

Note in particular, that the simple, yet previously unexpected,
form of the weak-field thermal Hall conductivity,
$\kappa_{xy} \sim T \sqrt{H}$, is now easily understood.
Summarizing the preceding discussion,
we write $\kappa_{xx} \sim C_{v} \ell$
and $\kappa_{xy} \sim \kappa_{xx} n_{v} \sigma_{\perp} \ell
\sim C_{v} \ell^{2} n_{v} \sigma_{\perp}$.  In the
weak-field limit, since quasiparticles are thermally excited,
$C_{v} \sim T^{2}$.  For small magnetic fields,
the mean free path is dominated by impurity
scattering.  Therefore
$\ell \sim \ell_{0} \sim \kappa_{e}(T)/C_{v} \sim 1/T$.
The vortex density is just proportional to the magnetic field,
$n_{v} \sim H$.  Our numerics yield
$\sigma_{\perp} \sim (k/p_{F}) R \sim T/\sqrt{H}$, which says that,
aside from being small by a factor of $k/p_{F}$, the skew cross section
is just proportional to the effective vortex radius, $R$.
Putting it all together yields
$\kappa_{xy} \sim (T^{2})(1/T)^{2}(H)(T/\sqrt{H}) \sim T\sqrt{H}$.
The simple form of this result is due to the simple nature of
the weak-field regime.  Here we have thermally excited
quasiparticles with a mean free path due to impurity scattering.
The only effect of the vortices is to add a small skew component
to the scattering.  The unusual $\sqrt{H}$ Hall response results
because, while the number of vortices goes like $H$, the skew cross
section per vortex goes like $1/\sqrt{H}$.

While our analysis appears to yield the correct functional form of
the thermal conductivity, we do not expect our results to
be quantitative due to the approximations we made in calculating
the single vortex cross section.
To obtain more quantitative results,
one must include the Berry phase contribution, consider anisotropic
Dirac nodes, and account (via empirical input) for the details of the
vortex core.  This is left for future work.

We are very grateful to Y. Zhang and N. P. Ong for sending us
their unpublished $\kappa_{xx}$ data and allowing us to reproduce
their $\kappa_{xy}$ plot in Fig.~\ref{fig:devfromscaling}.
We thank D. Huse, A. Millis, T. Senthil, and J. Ye for helpful discussions.
This work was supported by NSF Grant No. DMR-0201069.
A. V. was supported by an MIT Pappalardo Fellowship.


\begin{thebibliography}{99}
\bibitem{kappafootnote} Note that for an energy-dependent $\ell$, this
    simple relation between $\kappa_{xx}$ and $C_{v}$ is only approximate.
    We verified that while the more exact relation
    (see Ref.~\onlinecite{dur02}) modifies details, it preserves the qualitative
    form of our results.
\end{thebibliography}
\end{document}